\documentclass[amsmath,amssymb,showkeys,superscriptaddress,aps,prd,10pt,twocolumn]{revtex4-2}
\usepackage{graphicx}
\usepackage[utf8]{inputenc}
\usepackage[caption=false]{subfig}
\usepackage{multirow}
\usepackage{appendix}
\usepackage{graphicx,epstopdf}

\def\xslash{x\!\!\!\slash }

\def\vel{\left|}
\def\ver{\right|}
\usepackage{colortbl}
\usepackage{xcolor}
\usepackage[colorlinks=true, urlcolor=blue, linkcolor=green, citecolor=green]{hyperref}

\begin{document}

\title{Magnetic moments of the vector hidden-charmed tetraquark states}
\author{Ula\c{s} \"{O}zdem}%
\email[]{ulasozdem@aydin.edu.tr}
\affiliation{ Health Services Vocational School of Higher Education, Istanbul Aydin University, Sefakoy-Kucukcekmece, 34295 Istanbul, Turkey}

 
\begin{abstract}
The magnetic moments of the vector hidden-charmed tetraquark states that have been observed and can be expected to be observed experimentally have been determined using the light-cone sum rules taking into account the diquark-antidiquark structure with the quantum numbers $ J^{PC} =  1^{--}$  and $ J^{PC} =  1^{-+}$. 
Since these states are considered to have different flavors of light quarks, they have nonzero magnetic moments. The results obtained in this study can be checked for consistency by various methods. The magnetic moments of hadrons encompass useful knowledge about the distribution of charge and magnetization inside hadrons, which helps us to understand their geometrical shapes.
\end{abstract}

\maketitle

\section{Introduction} \label{1form}

Theoretically, the existence of states with a larger number of quarks besides baryons and mesons was proposed long ago. However, the first experimental discovery of these states occurred in 2003 with the observation of the X(3872) state by the Belle Collaboration \cite{Belle:2003nnu}. 
After the discovery of this particle, various experimental collaborations discovered many particles belonging to this new family that are still being discovered.
These newly discovered states not only arouse the interest of particle physicists, but also raise new questions about their inner structure and quantum numbers. Many models have been proposed to explain and decipher the nature of these states, and several studies have been conducted on them. 
However, their properties remain dubious, and their substructures and quantum numbers are also problematic. The properties of the reported tetraquark states have been interpreted differently in different studies. To resolve all these ambiguities, the properties of both the known and the newly observed states need to be further investigated. These studies could investigate complementary reactions or other decay modes for the currently known tetraquark states, or novel particles that may be observed can be investigated for their spectroscopic properties or possible decay modes to provide input for the experiments. Several interesting reviews provide detailed information on unconventional states, including a history of the subject and experimental and theoretical breakthroughs in recent years~\cite{Faccini:2012pj,Esposito:2014rxa,Chen:2016qju,Ali:2017jda,Esposito:2016noz,Olsen:2017bmm,Lebed:2016hpi,Guo:2017jvc,Nielsen:2009uh,Brambilla:2019esw,Liu:2019zoy, Agaev:2020zad, Dong:2021juy}.

Several vector hidden-charmed tetraquark states, such as Y(4220/4260), Y(4360/4390), Y(4630/4660) and so on, have been observed in recent years that cannot be well correlated in the standard meson with two quarks. The family of exotic vector states ($Y_{c \bar c}$ for short), called tetraquarks, contains at least four particles of hidden-charm with quantum numbers $J^{ PC }$ = $1^{--}$. In order to understand the nature of these states, many different models have been proposed and studies have been conducted on them (details of which can be found in the reviews~\cite{Chen:2016qju,Guo:2017jvc,Liu:2019zoy,Agaev:2020zad}). In Refs. \cite{Chen:2010ze,Wang:2013exa,Wang:2016mmg,Wang:2018rfw,Sundu:2018toi,Wang:2021qus}, the QCD sum rules have been also employed to explore the spectroscopic parameters of these states.  In Ref. \cite{Chen:2010ze}, a large number of interpolation currents were constructed for the $Y_{c \bar c}$ states, and the spectroscopic parameters of these states were studied using QCD sum rules with $J^{ PC }$ = $1^{++}$, $J^{ PC }$ = $1^{--}$, $J^{ PC }$ = $1^{-+}$  and $J^{ PC }$ = $1^{+-}$, and quark contents $[cq][\bar{c}\bar{q}]$ and $[cs][\bar{c}\bar{s}]$. It was taken into account that these states are in the diquark-antidiquark structure. While some of the obtained results are compatible with the experimentally discovered $Y_{c \bar c}$ states, they turned out to be incompatible with some of them. Moreover, some possible decay channels and the experimental search for these states are also discussed.  
In Ref. \cite{Wang:2013exa}, the mass and residue of the Y(4660) state were determined in the framework of the QCD sum rules. They found that $c\bar cs\bar s$ and $c\bar c(u\bar u+d \bar d)/\sqrt{2}$ diquark-antidiquark states favor the Y(4660) state with quantum numbers $J^{ PC }$ = $1^{--}$. They also excluded $c\bar cu\bar d$ diquark-antidiquark structure with quantum numbers $J^{ PC }$ = $1^{\pm -}$ for the Y(4360) state. In Ref. \cite{Wang:2016mmg}, they constructed different types of currents to interpolate both the vector and axial vector tetraquark states and obtain the spectroscopic parameters of the $Y_{c \bar c}$ states within the QCD sum rules. The numerical results support the assignment of the Y(4660) as a diquark-antidiquark type tetraquark state with quantum numbers $J^{ PC } = 1^{--}$. It has also been suggested that Y(4260) and Y(4360) may be mixed charmonium-tetraquark states. In Ref. \cite{Wang:2018rfw}, the tetraquark states of type $C \otimes \gamma_\mu C$ and $C \gamma_5 \otimes \gamma_5 \gamma_\mu C$ were constructed to calculate the mass and residue of the $Y_{c \bar c}$ states. Their analysis supported the assignment of Y(4660) and Y(4630) as vector tetraquark states of type $C \otimes \gamma_\mu C$ $c\bar c s \bar s$, assign Y(4360) and Y(4320) to the vector tetraquark state $c\bar c q \bar q$ of type $C \gamma_5 \otimes \gamma_5 \gamma_\mu C$ and do not assign Y(4260), Y(4220) and Y(4390) to the fixed vector tetraquark states.  In Ref. \cite{Sundu:2018toi}, the mass, decay constant, and strong decay channels of the Y(4660) state were evaluated by treating it as a bound state of a diquark and an antidiquark ($[cs][\bar{c}\bar{s}]$). It was shown that the results for the mass and total width of this state are in good agreement with the experimental data.   In Ref. \cite{Wang:2021qus}, they constructed the scalar, pseudoscalar, vector, axial vector, and tensor antidiquark states to obtain the mass spectrum of the vector tetraquark states with hidden charm via the QCD sum rules. Their predictions supported the identification of Y(4360), Y(4390), and Y(4660) as $[cq][\bar{c}\bar {q'}]$ vector tetraquark states with hidden charm with $J^{ PC } = 1^{--}$.

In addition to their spectroscopic properties, the electromagnetic form factors and multipole moments of hadrons can provide clues to their precise character, internal structure, and quantum numbers. We know that the electromagnetic multipole moments of hadrons, in particular their magnetic moments, which encompasses knowledge about the spatial distribution of charge and magnetization inside the hadrons, are related to the spatial distribution of quarks and gluons inside them. The study of the magnetic and higher multipole moments of hadrons is therefore attractive.
In this study, we compute the magnetic moments of $ Y_{c\bar c}$  states  in the diquark-antidiquark configuration  with the quantum numbers $ J^{PC} =  1^{--}$  and $ J^{PC} =  1^{-+}$ using the light-cone sum rule method \cite{Chernyak:1990ag, Braun:1988qv, Balitsky:1989ry}.  The  light-cone sum rule method  is based on the operator product expansion near the light-cone $x^2 \sim 0$ and parametrizes all the non-perturbative dynamics in the distribution amplitudes that have been used to treat many electromagnetic properties of conventional and non-conventional hadrons.

 This article is structured in the following manner. After the introduction in Sec. \ref{1form}, we present in Sec. \ref{2form} the formalism of the light-cone sum rule, which identifies the necessary tools to compute the magnetic moments of the $Y_{c\bar c}$ states. In Sec. \ref{3form}, we use the analytical formulas obtained in the previous section to perform numerical calculations of the magnetic moments and discuss the results.
 
 \begin{widetext}

 \section{Light-cone sum rule formalism for magnetic moments}\label{2form}
 
 In the light-cone sum rule technique, we compute a correlation function, which serves as the building block of the method, twice: once in terms of hadronic quantities such as coupling constants, form factors and electromagnetic multipole moments and second in terms of QCD parameters and photon distribution amplitudes available for different twists. The coefficients of the corresponding Lorentz structures from both representations of the correlation function are then equated and the quark-hadron duality approach is used to obtain the desired physical quantity.
 
As we have mentioned above, at the beginning of the analytic calculations of the magnetic moments it is necessary to write the correlation function, which plays an important role in the light-cone sum rules and is written as follows
\begin{equation}
 \label{edmn01}
\Pi _{\mu \nu }(p,q)=i\int d^{4}xe^{ip\cdot x}\langle 0|\mathcal{T}\{J_{\mu}^{i}(x)
J_{\nu }^{i \dagger }(0)\}|0\rangle_{\gamma}, 
\end{equation}%
where $\mathcal{T}$,  $J_{\mu}^i(x)$ and $\gamma$ represent the time-ordered product of two currents, the interpolating current of $Y_{c \bar c}$ states and the external electromagnetic field, respectively. We need explicit expressions for $J_{\mu}^i(x)$ to make progress in the calculations. In the diquark-antidiquark picture, $J_{\mu}^i(x)$ can be written in the following forms~\cite{Wang:2021qus}

\begin{eqnarray}
J^{1}_{\mu}(x)&=&\frac{\varepsilon\bar{\varepsilon}}{\sqrt{2}}\Big\{\big[u^{Tj}(x)Cc^k(x)\big] \big[ \bar{d}^m(x)\gamma_\mu C \bar{c}^{Tn}(x)\big]-\big[u^{Tj}(x)C\gamma_\mu c^k(x)\big]\big[\bar{d}^m(x)C \bar{c}^{Tn}(x) \big]\Big\},\nonumber\\
J^{2}_{\mu}(x)&=&\frac{\varepsilon\bar{\varepsilon}}{\sqrt{2}}\Big\{\big[u^{Tj}(x)Cc^k(x)\big]\big[ \bar{d}^m(x)\gamma_\mu  C \bar{c}^{Tn}(x)\big]+\big[u^{Tj}(x)C\gamma_\mu c^k(x)\big]\big[\bar{d}^m(x)C \bar{c}^{Tn}(x) \big] \Big\},\nonumber\\
J^{3}_{\mu}(x)&=&\frac{\varepsilon\bar{\varepsilon}}{\sqrt{2}}\Big\{\big[u^{Tj}(x)C\gamma_5c^k(x)\big]\big[ \bar{d}^m(x)\gamma_5\gamma_\mu C \bar{c}^{Tn}(x)\big]+\big[u^{Tj}(x)C\gamma_\mu\gamma_5 c^k(x)\big]\big[\bar{d}^m(x)\gamma_5C \bar{c}^{Tn}(x) \big] \Big\},\nonumber\\
J^{4}_{\mu}(x)&=&\frac{\varepsilon\bar{\varepsilon}}{\sqrt{2}}\Big\{\big[u^{Tj}(x)C\gamma_5c^k(x)\big]\big[ \bar{d}^m(x)\gamma_5\gamma_\mu C \bar{c}^{Tn}(x)\big]-\big[u^{Tj}(x)C\gamma_\mu\gamma_5 c^k(x)\big]\big[\bar{d}^m(x)\gamma_5C \bar{c}^{Tn}(x) \big] \Big\},\nonumber\\
%
J_{\mu}^{5}(x)&=&\frac{\varepsilon\bar{\varepsilon}}{\sqrt{2}}\Big\{\big[u^{Tj}(x)C\sigma_{\mu\nu} c^k(x)\big]\big[\bar{d}^m(x)\gamma^\nu C \bar{c}^{Tn}(x)\big]- \big[u^{Tj}(x)C\gamma^\nu c^k(x)\big] \big[\bar{d}^m(x)\sigma_{\mu\nu} C \bar{c}^{Tn}(x)\big] \Big\}, \nonumber\\
J_{\mu}^{6}(x)&=&\frac{\varepsilon\bar{\varepsilon}}{\sqrt{2}}\Big\{\big[u^{Tj}(x)C\sigma_{\mu\nu} c^k(x)\big]\big[\bar{d}^m(x)\gamma^\nu C \bar{c}^{Tn}(x)\big]+ \big[u^{Tj}(x)C\gamma^\nu c^k(x)\big] \big[\bar{d}^m(x)\sigma_{\mu\nu} C \bar{c}^{Tn}(x)\big] \Big\}, \nonumber\\
J_{\mu}^{7}(x)&=&\frac{\varepsilon\bar{\varepsilon}}{\sqrt{2}}\Big\{\big[u^{Tj}(x)C\sigma_{\mu\nu}\gamma_5 c^k(x)\big]\big[\bar{d}^m(x)\gamma_5\gamma^\nu C \bar{c}^{Tn}(x)\big]+ \big[u^{Tj}(x)C\gamma^\nu\gamma_5 c^k(x)\big]\big[\bar{d}^m(x)\gamma_5\sigma_{\mu\nu} C \bar{c}^{Tn}(x)\big] \Big\}  , \nonumber\\
J_{\mu}^{8}(x)&=&\frac{\varepsilon\bar{\varepsilon}}{\sqrt{2}}\Big\{\big[u^{Tj}(x)C\sigma_{\mu\nu}\gamma_5 c^k(x)\big]\big[\bar{d}^m(x)\gamma_5\gamma^\nu C \bar{c}^{Tn}(x)\big]- \big[u^{Tj}(x)C\gamma^\nu\gamma_5 c^k(x)\big]\big[\bar{d}^m(x)\gamma_5\sigma_{\mu\nu} C \bar{c}^{Tn}(x)\big] \Big\} , 
\end{eqnarray}
where $\varepsilon = \varepsilon^{ijk}$, $\bar{\varepsilon} = \varepsilon^{imn}$,  the i, j, k, m, n are color indices and the C is the charge conjugation matrix. Here the quantum numbers of $J^{1}_{\mu}$, $J^{3}_{\mu}$, $J^{5}_{\mu}$ and $J^{7}_{\mu}$ are the $J^{PC} = 1^{--}$, and the quantum numbers of $J^{2}_{\mu}$, $J^{4}_{\mu}$, $J^{6}_{\mu}$ and $J^{8}_{\mu}$ are the $J^{PC} = 1^{-+}$. At this point it should be noted that there are no experimentally observed vector hidden-charmed tetraquark states with the  quantum numbers $J^{PC} = 1^{-+}$.

 In the hadronic language, a complete set of hadronic states is inserted and the contributions of the lowest $Y_{c \bar c}$ states are separated to obtain the corresponding correlation function,
 
\begin{align}
\label{edmn04}
\Pi_{\mu\nu}^{Had} (p,q) = {\frac{\langle 0 \mid J_\mu (x) \mid
Y_{c\bar c}(p, \varepsilon^\theta) \rangle}{p^2 - m_{Y_{c\bar c}}^2}} \langle Y_{c\bar c}(p, \varepsilon^\theta) \mid Y_{c\bar c}(p+q, \varepsilon^\delta) \rangle_\gamma
\frac{\langle Y_{c\bar c}(p+q,\varepsilon^\delta) \mid {J_\nu^{ \dagger}} (0) \mid 0 \rangle}{(p+q)^2 - m_{Y_{c\bar c}}^2} + \cdots,
\end{align}
where  dots denote the effects of the higher states and continuum. The matrix elements in Eq. (\ref{edmn04})
are expressed as
\begin{align}
\label{edmn05}
\langle Y_{c\bar c}(p+q,\varepsilon^\delta) \mid {J_\nu^{ \dagger}} (0) \mid 0 \rangle&=\lambda_{Y_{c\bar c}} \varepsilon_\nu^\delta\,,
\\
\nonumber\\
\langle 0 \mid J_\mu(x) \mid Y_{c\bar c}(p,\varepsilon^\theta) \rangle &= \lambda_{Y_{c\bar c}} \varepsilon_\mu^\theta\,,
\\
\nonumber\\
\langle Y_{c\bar c}(p,\varepsilon^\theta) \mid  Y_{c\bar c} (p+q,\varepsilon^{\delta})\rangle_\gamma &= - \varepsilon^\tau (\varepsilon^{\theta})^\alpha (\varepsilon^{\delta})^\beta \Big\{ G_1(Q^2)~ (2p+q)_\tau ~g_{\alpha\beta}  + G_2(Q^2)~ ( g_{\tau\beta}~ q_\alpha -  g_{\tau\alpha}~ q_\beta) \nonumber\\ &- \frac{1}{2 m_{Y_{c\bar c}}^2} G_3(Q^2)~ (2p+q)_\tau ~q_\alpha q_\beta  \Big\},\label{edmn06}
\end{align}
where $\varepsilon^\tau$ is polarization of the photon,      $\lambda_{Y_{c\bar c}}$ is residue of the $Y_{c\bar c}$  states and   $G_i(Q^2)$'s are electromagnetic form factors,  with  $Q^2=-q^2$.
To calculate the magnetic moment, we need only $G_2(Q^2)$ of the form factors described above. 
The magnetic form factor, $F_M(Q^2)$, is written as follows
\begin{align}
\label{edmn07}
&F_M(Q^2) = G_2(Q^2)\,.
\end{align} 

Using Eqs. (\ref{edmn04})-(\ref{edmn06}) and after doing some necessary calculations the final form of the correlation function is  obtained as
\begin{align}
\label{edmn09}
 \Pi_{\mu\nu}^{Had}(p,q) &=  \frac{\varepsilon_\rho \, \lambda_{Y_{c\bar c}}^2}{ [m_{Y_{c\bar c}}^2 - (p+q)^2][m_{Y_{c\bar c}}^2 - p^2]}
 \Big\{ G_2 (Q^2) \Big(q_\mu g_{\rho\nu} - q_\nu g_{\rho\mu} -
\frac{p_\nu}{m_{Y_{c\bar c}}^2}  \big(q_\mu p_\rho - \frac{1}{2}
Q^2 g_{\mu\rho}\big) 
 + \nonumber\\
 &  +
\frac{(p+q)_\mu}{m_{Y_{c\bar c}}^2}  \big(q_\nu (p+q)_\rho+ \frac{1}{2}
Q^2 g_{\nu\rho}\big)
-  
\frac{(p+q)_\mu p_\nu p_\rho}{m_{Y_{c\bar c}}^4} \, Q^2
\Big)
\nonumber\\
&
+\mbox{other independent structures}\Big\}\,+\cdots.
\end{align}

 The $F_M(Q^2=0)$ is proportional to the magnetic moment $\mu_{Y_{c\bar c}}$:
\begin{align}
\label{edmn08}
&\mu_{Y_{c\bar c}} = \frac{ e}{2\, m_{Y_{c\bar c}}} \,F_M(0).
\end{align}

The correlation function is determined in terms of the QCD degrees of freedom and the photon distribution amplitudes in the second step of the calculation of the magnetic moment of $Y_{c\bar c}$ states. In the QCD representation, we use Wick's theorem to contract the corresponding quark fields after replacing the explicit expressions of the interpolating currents in the correlation function. For instance, the result for the current $J_{\mu}^1$ is as follows:

\begin{eqnarray}
\Pi _{\mu \nu }^{\mathrm{QCD}}(p,q)&=&i\frac{\varepsilon \bar{\varepsilon}\varepsilon^\prime \bar{\varepsilon}^\prime}{2}\int d^{4}xe^{ip\cdot x} \langle 0 \mid \Bigg\{  \mathrm{%
Tr}\Big[  S_{c}^{kk^{\prime }}(x)\widetilde S_{u}^{jj^{\prime }}(x)\Big]    
\mathrm{Tr}\Big[ \gamma _{\mu } \widetilde S_{c}^{n^{\prime
}n}(-x)\gamma _{\nu }S_{d}^{m^{\prime }m}(-x)\Big]\notag \\
&&- \mathrm{%
Tr}\Big[ S_{c}^{kk^{\prime }}(x)\gamma
_{\nu}\widetilde S_{u}^{jj^{\prime }}(x)\Big]    
\mathrm{Tr}\Big[ \gamma _{\mu } \widetilde S_{c}^{n^{\prime
}n}(-x)S_{d}^{m^{\prime }m}(-x)\Big]\notag \\
&& -\mathrm{Tr}\Big[  \gamma _{\mu} S_{c}^{kk^{\prime }}(x)\widetilde S_{u}^{jj^{\prime }}(x)\Big] \mathrm{Tr}\Big[ \widetilde S_{c}^{n^{\prime
}n}(-x)\gamma _{\nu}S_{d}^{m^{\prime }m}(-x)\Big]\notag\\
&&+\mathrm{Tr}\Big[  \gamma _{\mu } S_{c}^{kk^{\prime }}(x) \gamma _{\nu}\widetilde S_{u}^{jj^{\prime }}(x)\Big] \mathrm{Tr}\Big[\widetilde S_{c}^{n^{\prime
}n}(-x)S_{d}^{m^{\prime }m}(-x)\Big]\Bigg\} \mid 0 \rangle_{\gamma} ,  \label{QCDSide1}
\end{eqnarray}%
where%
\begin{equation*}
\widetilde{S}_{c(q)}^{ij}(x)=CS_{c(q)}^{ij\mathrm{T}}(x)C,
\end{equation*}%
with $S_{q(c)}(x)$ being the quark propagators.
In the $x$-space for the light-quark propagator we use in the $m_q\rightarrow 0$ limit
\begin{align}
\label{edmn12}
S_{q}(x)&=i \frac{{\xslash}}{2\pi ^{2}x^{4}} 
- \frac{\langle \bar qq \rangle }{12} 
- \frac{ \langle \bar qq \rangle }{192}m_0^2 x^2 
-\frac {i g_s }{32 \pi^2 x^2} ~G^{\mu \nu} (x) \Big[\rlap/{x} 
\sigma_{\mu \nu} +  \sigma_{\mu \nu} \rlap/{x}
 \Big],
\end{align}
where $\langle \bar qq \rangle$ is light quark  condensate, $m_0^2$ is defined via the relation  $\langle 0 \mid \bar  q\, g_s\, \sigma_{\alpha\beta}\, G^{\alpha\beta}\, q \mid 0 \rangle = m_0^2 \,\langle \bar qq \rangle $. 

The charm-quark propagator is given, in association with the second kind Bessel functions $K_{i}(x)$, as
\begin{align}
\label{edmn13}
S_{c}(x)&=\frac{m_{c}^{2}}{4 \pi^{2}} \Bigg[ \frac{K_{1}\Big(m_{c}\sqrt{-x^{2}}\Big) }{\sqrt{-x^{2}}}
+i\frac{{\xslash}~K_{2}\Big( m_{c}\sqrt{-x^{2}}\Big)}
{(\sqrt{-x^{2}})^{2}}\Bigg]
-\frac{g_{s}m_{c}}{16\pi ^{2}} \int_0^1 dv\, G^{\mu \nu }(vx)\Bigg[ \big(\sigma _{\mu \nu }{\xslash}
  +{\xslash}\sigma _{\mu \nu }\big)\nonumber\\
  &\times \frac{K_{1}\Big( m_{c}\sqrt{-x^{2}}\Big) }{\sqrt{-x^{2}}}
+2\sigma_{\mu \nu }K_{0}\Big( m_{c}\sqrt{-x^{2}}\Big)\Bigg].
\end{align}%
 where $v$  is line variable and  $G^{\mu\nu}$ is the gluon field strength tensor.   The perturbative or free component of the propagators of the light and heavy quarks corresponds to the first term, while the remainder belongs to the interacting parts (with gluon or
QCD vacuum) as nonperturbative contributions.

The correlation function in Eq. (\ref{QCDSide1}) includes  different types of contributions: the photon can be emitted both perturbatively or
non-perturbatively. In first case, one of the free light or heavy quark 
propagators in Eq.~(\ref{QCDSide1}) is replaced by
\begin{align}
\label{sfree}
S^{free} \rightarrow \int d^4y\, S^{free} (x-y)\,\rlap/{\!A}(y)\, S^{free} (y)\,,
\end{align}
the remaining propagators are replaced with the full quark propagators. The light-cone sum rule analyses are most conveniently done in the fixed-point gauge. The most important advantage of a fixed-point gauge is that the external field is expressed as being  associated with the field strength tensor. For the electromagnetic field,
it is defined by $x_\mu A^\mu =0$. 
In this gauge, the external electromagnetic potential is given by 
\begin{align}
\label{AAA}
 &A_\alpha = -\frac{1}{2} F_{\alpha\beta}y^\beta 
   = -\frac{1}{2} (\varepsilon_\alpha q_\beta-\varepsilon_\beta q_\alpha)\,y^\beta.
\end{align}
Equation (\ref{AAA}) is plugged into Eq. (\ref{sfree}), as a result of which we obtain
 \begin{align}
  S^{free} \rightarrow -\frac{1}{2} (\varepsilon_\alpha q_\beta-\varepsilon_\beta q_\alpha)
  \int\, d^4y \,y^{\beta}\, 
  S^{free} (x-y)\,\gamma_{\alpha}\,S^{free} (y)\,,
 \end{align}

After some lengthy calculations for $S_q^{free}$ and $S_c^{free}$, we obtain their final form as follows:

\begin{eqnarray}
&& S_q^{free}=\frac{e_q}{32 \pi^2 x^2}\Big(\varepsilon_\alpha q_\beta-\varepsilon_\beta q_\alpha\Big)
 \Big(\xslash\sigma_{\alpha \beta}+\sigma_{\alpha\beta}\xslash\Big),\nonumber\\
&& S_c^{free}=-i\frac{e_c m_c}{32 \pi^2}
\Big(\varepsilon_\alpha q_\beta-\varepsilon_\beta q_\alpha\Big)
\Big[2\sigma_{\alpha\beta}K_{0}\Big( m_{c}\sqrt{-x^{2}}\Big)
 +\frac{K_{1}\Big( m_{c}\sqrt{-x^{2}}\Big) }{\sqrt{-x^{2}}}
 \Big(\xslash\sigma_{\alpha \beta}+\sigma_{\alpha\beta}\xslash\Big)\Big].
\end{eqnarray}

In the second case one of the light quark 
propagators in Eq.~(\ref{QCDSide1}) is replaced by
\begin{align}
\label{neweq}
S_{\alpha\beta}^{ab} \rightarrow -\frac{1}{4} (\bar{q}^a \Gamma_i q^b)(\Gamma_i)_{\alpha\beta},
\end{align}
and the remaining propagators are full quark propagators including 
the perturbative as well as the nonperturbative contributions. Here as an example, we give a short detail of the calculations of the QCD representations.
 In second case for simplicity, we only consider the first trace in Eq.~(\ref{QCDSide1}),
\begin{eqnarray}\label{QCDES}
\Pi _{\mu \nu }^{\mathrm{QCD}}(p,q)&=&i\frac{\varepsilon \bar{\varepsilon}\varepsilon^\prime \bar{\varepsilon}^\prime}{2}\int d^{4}xe^{ip\cdot x} \langle 0 \mid \Bigg\{  \mathrm{%
Tr}\Big[  S_{c}^{kk^{\prime }}(x)\Gamma_i\Big]    
\mathrm{Tr}\Big[ \gamma _{\mu } \widetilde S_{c}^{n^{\prime
}n}(-x)\gamma _{\nu }S_{d}^{m^{\prime }m}(-x)\Big]\nonumber\\
&&+\mathrm{%
Tr}\Big[  S_{c}^{kk^{\prime }}(x) \widetilde S_u^{jj'}(x)\Big]    
\mathrm{Tr}\Big[ \gamma _{\mu } \widetilde S_{c}^{n^{\prime
}n}(-x)\gamma _{\nu }\Gamma_i\Big]
\Bigg\} \mid 0 \rangle_{\gamma}+.... , 
\end{eqnarray}%
where $\Gamma_i = I, \gamma_5, \gamma_\mu, i\gamma_5 \gamma_\mu, \sigma_{\mu\nu}/2$.
 
 By replacing one of light propagators with the expressions in Eq. (\ref{edmn13})
 and making use of 
 \begin{align}
  \bar q^a(x)\Gamma_i q^{a'}(0)\rightarrow \frac{1}{3}\delta^{aa'}\bar q(x)\Gamma_i q(0),
 \end{align}
 Eq. (\ref{QCDES}) takes the form
 \begin{eqnarray}
\label{QCDES2}
\Pi _{\mu \nu }^{\mathrm{QCD}}(p,q)&=&i\frac{\varepsilon \bar{\varepsilon}\varepsilon^\prime \bar{\varepsilon}^\prime}{2}\int d^{4}xe^{ip\cdot x} \Bigg\{  \mathrm{%
Tr}\Big[  S_{c}^{kk^{\prime }}(x)\Gamma_i\Big]    
\mathrm{Tr}\Big[ \gamma _{\mu } \widetilde S_{c}^{n^{\prime
}n}(-x)\gamma _{\nu }S_{d}^{m^{\prime }m}(-x)\Big] \delta^{jj'}\nonumber\\
&&+\mathrm{%
Tr}\Big[  S_{c}^{kk^{\prime }}(x) \widetilde S_u^{jj'}(x)\Big]    
\mathrm{Tr}\Big[ \gamma _{\mu } \widetilde S_{c}^{n^{\prime
}n}(-x)\gamma _{\nu }\Gamma_i\Big] \delta^{m'm}
\Bigg\} \frac{1}{12} \langle \gamma(q) |\bar q(x)\Gamma_i q(0)|0\rangle 
+....
 \end{eqnarray}

 Similarly, when a light propagator interacts with the photon, 
a gluon may be released from one of the remaining three propagators. 
The expression obtained in this case is as follows:
\begin{eqnarray}
\label{QCDES4}
\Pi _{\mu \nu }^{\mathrm{QCD}}(p,q)&=&i\frac{\varepsilon \bar{\varepsilon}\varepsilon^\prime \bar{\varepsilon}^\prime}{2}\int d^{4}xe^{ip\cdot x}  \Bigg\{  \mathrm{%
Tr}\Big[  S_{c}^{kk^{\prime }}(x)\Gamma_i\Big]    
\mathrm{Tr}\Big[ \gamma _{\mu } \widetilde S_{c}^{n^{\prime
}n}(-x)\gamma _{\nu }S_{d}^{m^{\prime }m}(-x)\Big] \Big[\Big(\delta^{kj}\delta^{k'j'}-\frac{1}{3}\delta^{kk'}\delta^{jj'}\Big)\nonumber\\
&&+\Big(\delta^{jn'}\delta^{j'n}-\frac{1}{3}\delta^{n'n}\delta^{jj'}\Big)
+\Big(\delta^{jm'}\delta^{j'm}-\frac{1}{3}\delta^{m'm}\delta^{jj'}\Big)\Big]
\nonumber\\
&&+\mathrm{%
Tr}\Big[  S_{c}^{kk^{\prime }}(x) \widetilde S_u^{jj'}(x)\Big]    
\mathrm{Tr}\Big[ \gamma _{\mu } \widetilde S_{c}^{n^{\prime
}n}(-x)\gamma _{\nu }\Gamma_i\Big] \Big[ \Big(\delta^{km'} \delta^{k'm}-\frac{1}{3} \delta^{m'm}\delta^{kk'}\Big)\nonumber\\
&&+\Big(\delta^{jm'} \delta^{j'm}-\frac{1}{3} \delta^{m'm}\delta^{jj'}\Big)
+\Big(\delta^{n'm'} \delta^{nm}-\frac{1}{3} \delta^{m'm}\delta^{n'n}\Big)
\Bigg\}
\nonumber\\
&& \times \frac{1}{32} \langle \gamma(q) |\bar q(x)\Gamma_i G_{\mu\nu}(vx) q(0)|0\rangle 
+...,
 \end{eqnarray}
where we inserted
\begin{align}
\label{QCDES5}
 \bar q^a(x)\Gamma_i G_{\mu\nu}^{bb'}(vx) q^{a'}(0)\rightarrow \frac{1}{8}\Big(\delta^{ab}\delta^{a'b'}
 -\frac{1}{3}\delta^{aa'}\delta^{bb'}\Big)\bar q(x)\Gamma_i G_{\mu\nu}(vx) q(0).
\end{align}

As can be seen, matrix elements representing non-perturbative contributions such as $\langle \gamma(q)\vel \bar{q}(x) \Gamma_i q(0) \ver 0\rangle$
and $\langle \gamma(q)\vel \bar{q}(x) \Gamma_i G_{\mu\nu}(vx)q(0) \ver 0\rangle$ appear. 
These matrix elements can be expressed associated with photon distribution amplitudes (DAs) and wave functions with definite twists, whose expressions are borrowed from  Ref.~\cite{Ball:2002ps}. Besides these matrix elements non-local operators such as two gluons ($\bar q G G q$) and  four quarks ($\bar qq \bar q q$) are expected to seem. However it is known that the effects of such operators are small, which is justified by the conformal spin expansion \cite{Balitsky:1987bk,Braun:1989iv}, and thus we shall ignore them. The QCD representation of the correlation function is obtained by using Eqs.~(\ref {QCDSide1}-\ref {QCDES5}).  Then, the Fourier transformation
is applied to transfer expressions in x-space to the momentum space.

In conclusion, the structure $q_\mu \varepsilon_\nu$ is chosen from both representations and the coefficients of the structure are matched in both hadronic and QCD representations. Then, Borel transformation and continuum subtraction are used to suppress the effects of the continuum and higher states. These steps are routine and tedious in the light-cone sum rule method, so we will not discuss them in detail here. Technical details on these applications can be found in Ref.\cite{Azizi:2018duk}. Thus, as an example for the current $J_{\mu}^1$, the light-cone sum rule for $Y_{c \bar c}$ states is as follows:
\begin{align}
 &\mu_{Y_{c\bar c}}\,\, \lambda_{Y_{c\bar c}}^2  = e^{\frac{m_{Y_{c\bar c}}^2}{M^2}} \,\, \Delta_1 (M^2,s_0),
\end{align}
where the explicit expression of the $\Delta_1 (M^2,s_0)$ function is  presented in the Appendix.  
The analytic calculations of the magnetic moments of $Y_{c\bar c}$ states come to an end here. In the following section we will use these analytical results to perform numerical calculations.


\end{widetext}

\section{Numerical analysis and conclusions} \label{3form}

We assume the following parameters to perform the numerical calculations for the magnetic moments of the $Y_{c\bar c}$ states. The masses of the light quarks are $m_u=m_d=0$, the mass of the c-quark is $m_c = (1.275\pm 0.025)\,$GeV, the condensates of the light quarks are $\langle \bar uu\rangle $
=$\langle \bar dd \rangle $=$(-0.24\pm0.01)^3\,$GeV$^3$~\cite{Ioffe:2005ym}, the gluon condensate is $\langle g_s^2G^2\rangle = 0.88~ $GeV$^4$~\cite{Nielsen:2009uh} and  the magnetic susceptibility $\chi=-2.85 \pm 0.5~\mbox{GeV}^{-2}$~\cite{Rohrwild:2007yt}. To progress numerical analysis of the magnetic moment of these $Y_{c\bar c}$ states, numerical values of the mass and residue parameters of these $Y_{c\bar c}$ states are also required. These values have been computed in Ref.~\cite{Wang:2021qus} using mass sum rules which are presented in Table \ref{massres}. 
The wave functions in the distribution amplitudes of the photon and all necessary expressions about these functions are taken from Ref.~\cite{Ball:2002ps}.

\begin{table}[htp]
	\addtolength{\tabcolsep}{10pt}
	\caption{Mass and the residue values of the $Y_{c\bar c}$ states which are borrowed from Ref. \cite{Wang:2021qus}.}
	\label{massres}
		\begin{center}
\begin{tabular}{l|c|cccc}
	   \hline\hline
   $Y_{c\bar c}$ State&  $m_{Y_{c\bar c}}$~\mbox{[GeV]}&$\lambda_{Y_{c\bar c}}$($\times 10^{-2}$)~\mbox{[GeV$^5$] }\\
\hline\hline
$J^1_\mu$&   $ 4.66 \pm 0.07$        &  $ 7.19 \pm 0.84$\\
$J^2_\mu$&   $ 4.61 \pm 0.07$        &  $ 6.69 \pm 0.80 $\\
$J^3_\mu$ &  $ 4.35 \pm 0.08$        &  $ 4.32 \pm 0.61 $\\
$J^4_\mu$ &  $ 4.66 \pm 0.09$        &  $ 6.67 \pm 0.82 $\\
$J^5_\mu$ &  $ 4.53 \pm 0.07$        &  $ 10.3 \pm 1.40 $\\
$J^6_\mu$ &  $ 4.65 \pm 0.08$        &  $ 11.3 \pm 1.50 $\\
$J^7_\mu$ &  $ 4.48 \pm 0.08$        &  $ 9.47 \pm 1.27 $\\
$J^8_\mu$ &  $ 4.55 \pm 0.07$        &  $ 10.6 \pm 1.40 $\\
	   \hline\hline
\end{tabular}
\end{center}
\end{table}

\begin{widetext}
 
\begin{figure}[htp]
\centering
\subfloat[]{\includegraphics[width=0.5\textwidth]{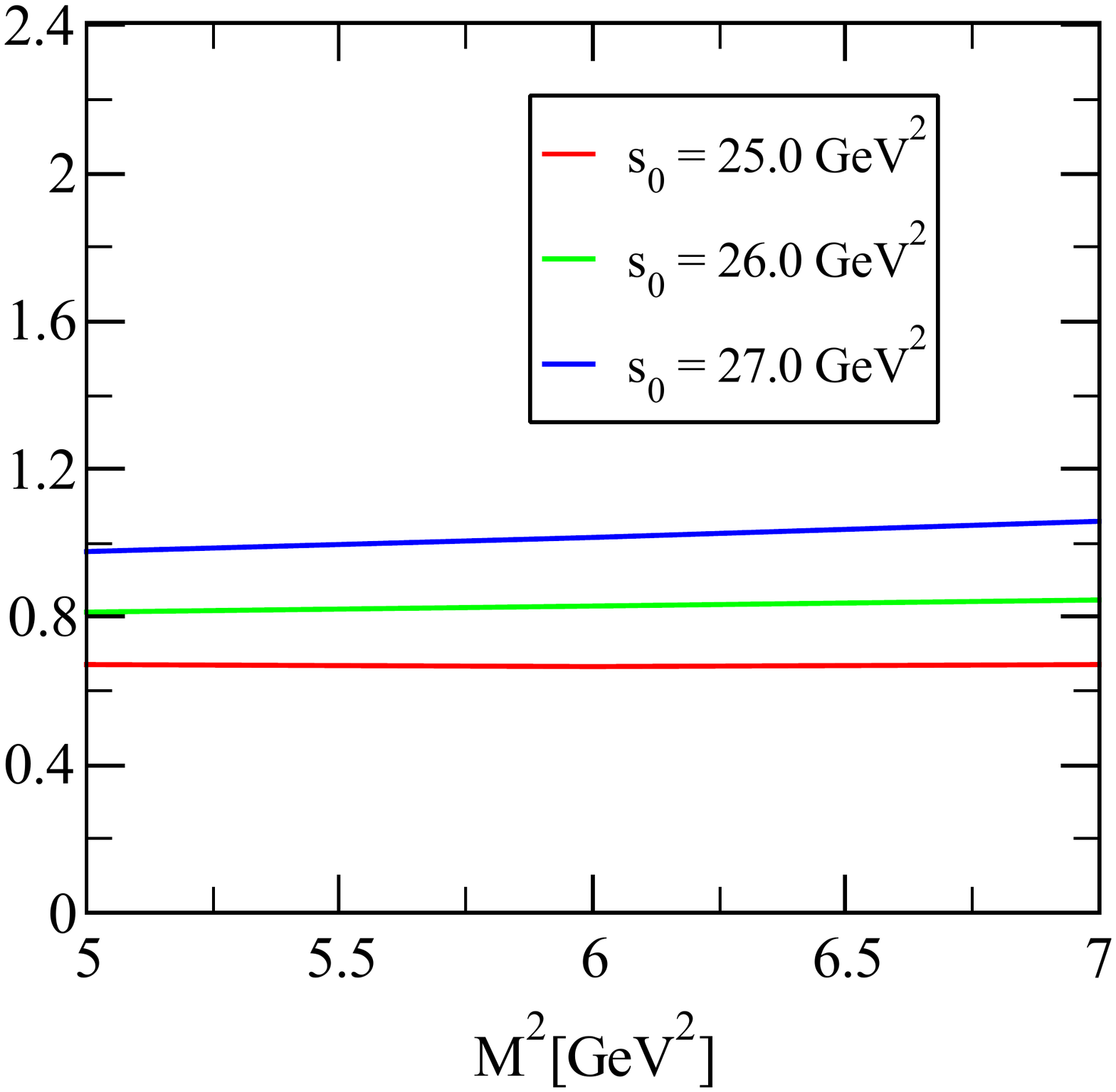}}
\subfloat[]{\includegraphics[width=0.5\textwidth]{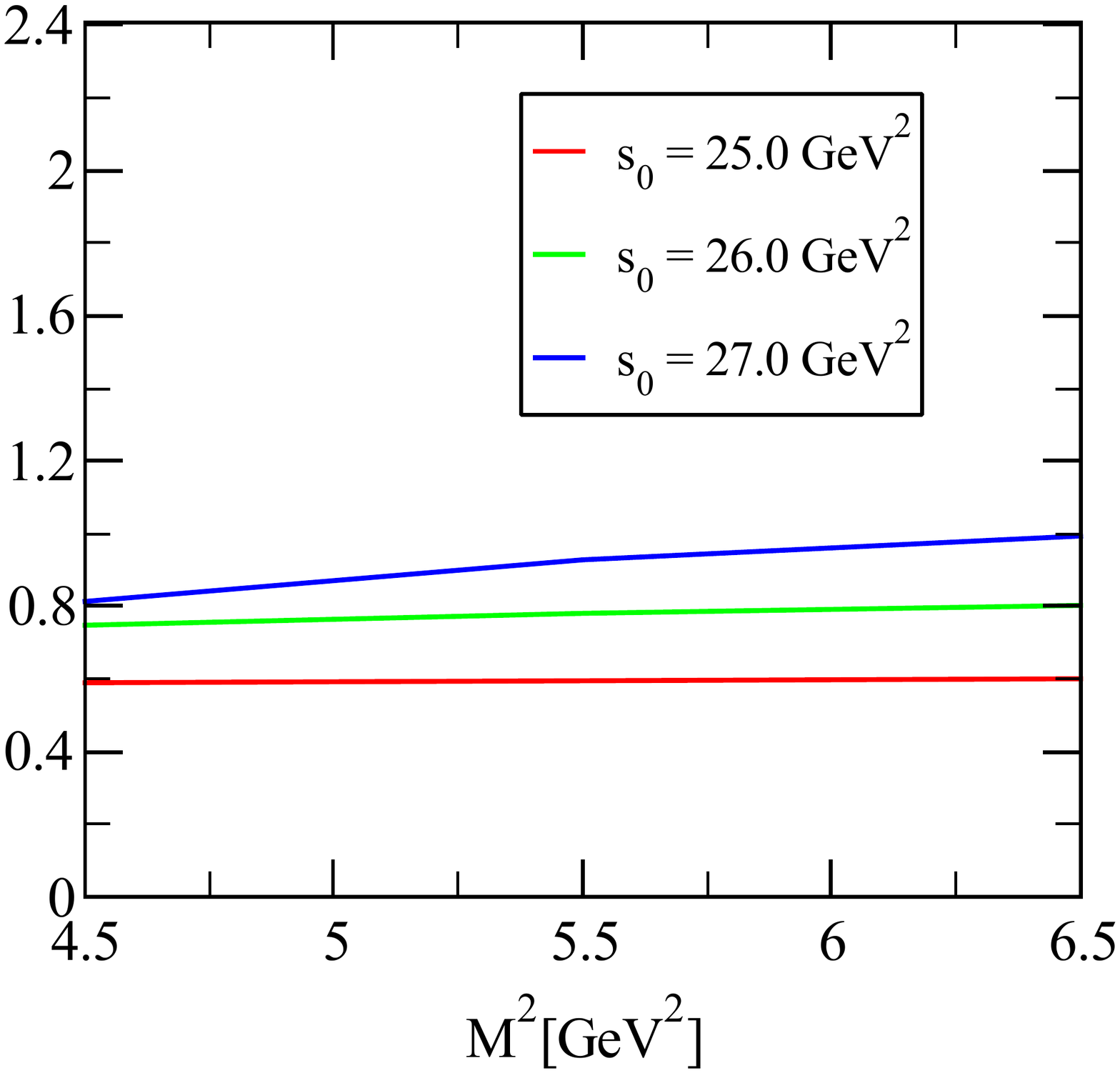}}\\
\subfloat[]{\includegraphics[width=0.5\textwidth]{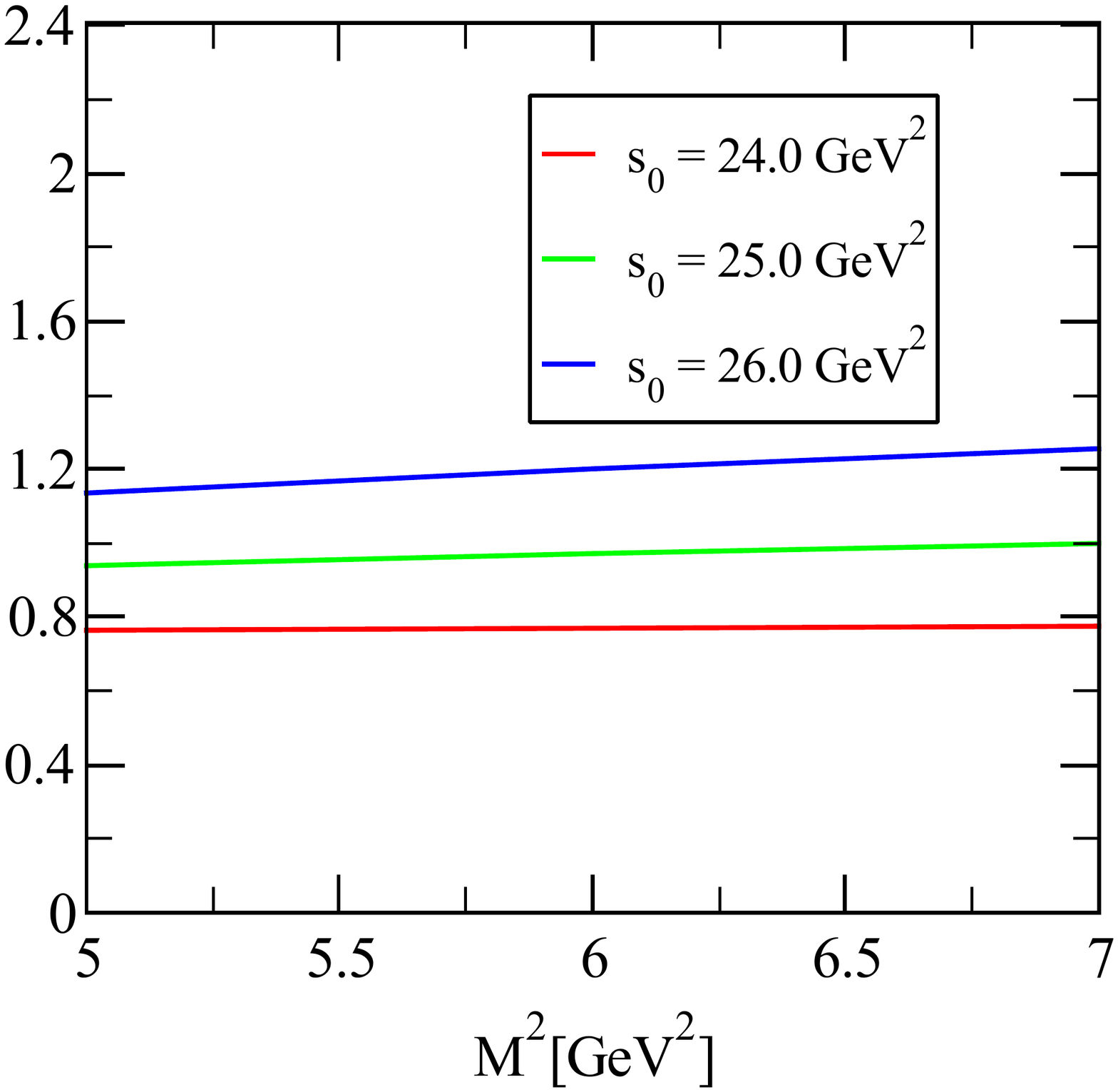}}
\subfloat[]{\includegraphics[width=0.5\textwidth]{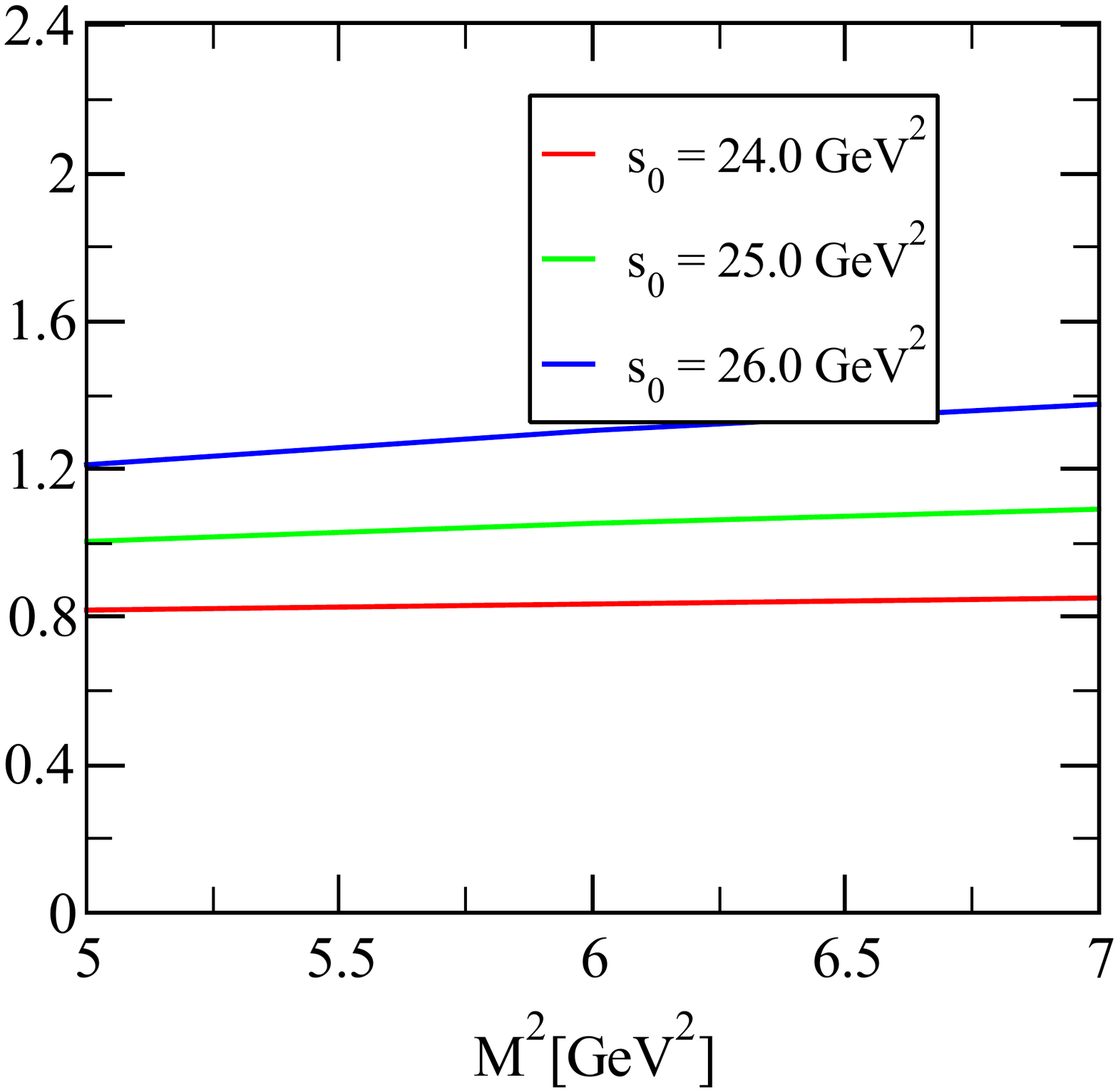}}
 \caption{The magnetic moments  versus $M^2$ at three fixed values of $s_0$; (a), (b), (c) and (d) for $J^{1}_{\mu}$, $J^{3}_{\mu}$, $J^{5}_{\mu}$ and $J^{7}_{\mu}$ states, respectively  (in unit of $\mu_N$ ).}
 \label{Msqfig1}
  \end{figure}
  
  \end{widetext}

\begin{widetext}
 
\begin{figure}[htp]
\centering
\subfloat[]{\includegraphics[width=0.5\textwidth]{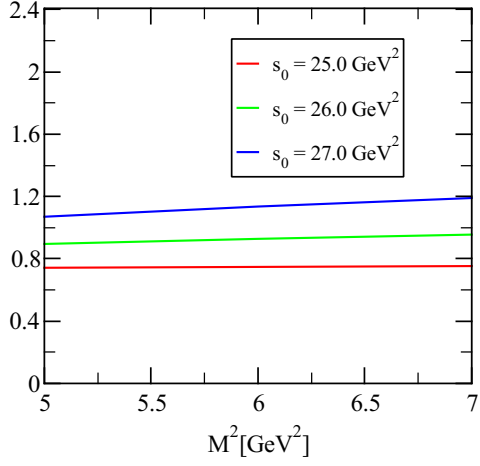}}
\subfloat[]{\includegraphics[width=0.5\textwidth]{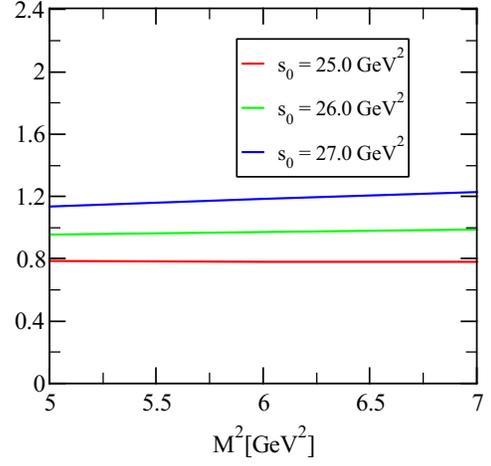}}\\
\subfloat[]{\includegraphics[width=0.5\textwidth]{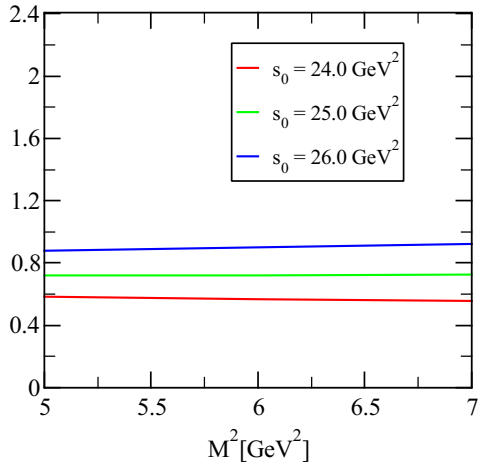}}
\subfloat[]{\includegraphics[width=0.5\textwidth]{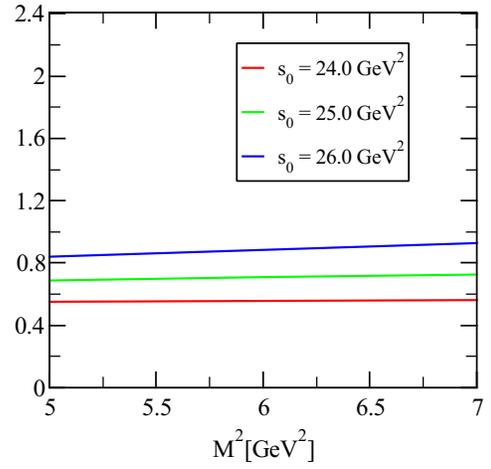}}
 \caption{The magnetic moments  versus $M^2$ at three fixed values of $s_0$; (a), (b), (c) and (d) for $J^{2}_{\mu}$, $J^{4}_{\mu}$, $J^{6}_{\mu}$ and $J^{8}_{\mu}$ states, respectively  (in unit of $\mu_N$ ).}
 \label{Msqfig2}
  \end{figure}
  
  \end{widetext}

In addition to the above input parameters, the light-cone sum rule method includes two other arbitrary parameters, the Borel mass ($M^2$) and the continuum threshold ($s_0$). According to the philosophy of the method, the physical quantity under study should be independent of the variation of these parameters.  To achieve this, we need to add some physical constraints, such as the convergence of the operator product expansion (OPE) and the pole contribution (PC). This means that the edges of the working windows for these arbitrary parameters should be set by the convergence of the OPE and the constraint on the PC.
We use two-criteria to determine the working region of $M^2$: the lower bound of $M^2$ is constrained  by the OPE convergence, demanding the higher twist and higher condensates terms to be less than $10 \%$ of the total. The upper bound of $M^2$ is constrained by the PC
\begin{align}
 \mbox{PC} =\frac{\Delta_1 (M^2,s_0)}{\Delta_1 (M^2,\infty)}\geq 30\%.
\end{align}
The continuum threshold $s_0$ is not arbitrary and it is related to the energy of the first excited state in the initial channel. However, since we have very limited information on the energy of excited states, we should decide how to choose working interval of the $s_0$.
Analysis of various sum rules predicted that  $s_0 \simeq (m_{ground}+0.5^{+0.2}_{-0.2})^2$~GeV$^{2}$. For more precise determination of continuum threshold, we impose the dominance of PC and OPE convergence limitations. As a results of these limitations, for $Y_{c\bar c}$ states we choose $s_0 \simeq (m_{Y_{c\bar c}}+0.5^{+0.1}_{-0.1})^2$~GeV$^{2}$.
Due to the above constraints, the following working windows for these arbitrary parameters together with PC and OPE convergence are shown in the Table \ref{parameter}. In Figs. \ref{Msqfig1} and \ref{Msqfig2}, we show the dependencies of the magnetic moments versus $M^2$ at three fixed values of $s_0$. As you can see from these figures, the variation of magnetic moments with respect to $M^2$ is quite stable. Although the variation is high compared to $s_0$, this variation remains within the errors of the method used.

\begin{widetext}

\begin{table}[htp]
	\addtolength{\tabcolsep}{10pt}
	\caption{Working regions of the Borel mass parameters, continuum threshold, PC and OPE convergence for magnetic moments.}
	\label{parameter}
		\begin{center}
\begin{tabular}{l|c|c|c|cc}
	   \hline\hline
   $Y_{c\bar c}$ State&  $s_0$~\mbox{[GeV$^2$]}&$M^2$~\mbox{[GeV$^2$]}& \mbox{PC (average)} & \mbox{OPE}\\
\hline\hline
$J^1_\mu$&   $ 25.0-27.0$        &  $ 5.0-7.0 $&$46 \%$& $<2\%$\\
$J^2_\mu$&   $ 25.0-27.0$        &  $ 5.0-7.0 $&$45 \%$&$<3\%$\\
$J^3_\mu$ &  $ 22.0-24.0$        &  $ 4.5-6.5 $&$43 \%$&$<2\%$\\
$J^4_\mu$ &  $ 25.0-27.0$        &  $ 5.0-7.0 $&$44 \%$&$<3\%$\\
$J^5_\mu$ &  $ 24.0-26.0$        &  $ 5.0-7.0 $&$43 \%$&$<2\%$\\
$J^6_\mu$ &  $ 24.0-26.0$        &  $ 5.0-7.0 $&$43 \%$&$<3\%$\\
$J^7_\mu$ &  $ 24.0-26.0$        &  $ 5.0-7.0 $&$45 \%$&$<2\%$\\
$J^8_\mu$ &  $ 24.0-26.0$        &  $ 5.0-7.0 $&$44 \%$&$<2\%$\\
	   \hline\hline
\end{tabular}
\end{center}
\end{table}

\end{widetext}

The magnetic moment results obtained for the $Y_{c\bar c}$ states with these input parameters are given in Table \ref{magmom} after determining all the necessary input parameters for the numerical analysis, both in its natural unit  ($\frac{ e}{2\, m_{Y_{c\bar c}}}$) and in the nuclear magneton unit ($\mu_N = \frac{ m_N}{ m_{Y_{c\bar c}}}$).
\begin{table}[htp]
	\addtolength{\tabcolsep}{10pt}
	\caption{Magnetic moments of $Y_{c\bar c}$ states.}
	\label{magmom}
		\begin{center}
\begin{tabular}{l|c|c|ccc}
	   \hline\hline
   $Y_{c\bar c}$ State &$J^{PC}$ & $\mu$~\mbox{\big[$\frac{ e}{2\, m_{Y_{c\bar c}}}$\big]}& $\mu$~\mbox{$\big[ \mu_N\big]$}\\
\hline\hline
$J^1_\mu$ &   $ 1^{--}$   &  $ 4.24^{+1.01}_{-0.91} $     &  $ 0.85^{+0.20}_{-0.18} $\\
$J^2_\mu$ &   $ 1^{-+}$   &  $ 4.68^{+1.17}_{-1.04} $     &  $ 0.95^{+0.23}_{-0.21} $\\
$J^3_\mu$ &   $ 1^{--}$   &  $ 3.74^{+1.13}_{-1.00} $     &  $ 0.80^{+0.25}_{-0.21} $\\
$J^4_\mu$ &   $ 1^{-+}$   &  $ 4.94^{+1.16}_{-1.03} $     &  $ 1.00^{+0.22}_{-0.21} $\\
$J^5_\mu$ &   $ 1^{--}$   &  $ 4.81^{+1.26}_{-1.12} $     &  $ 1.00^{+0.26}_{-0.24} $\\
$J^6_\mu$ &   $ 1^{-+}$   &  $ 3.68^{+0.88}_{-0.79} $     &  $ 0.74^{+0.18}_{-0.16} $\\
$J^7_\mu$ &   $ 1^{--}$   &  $ 5.18^{+1.40}_{-1.26} $     &  $ 1.08^{+0.30}_{-0.26} $\\
$J^8_\mu$ &   $ 1^{-+}$   &  $ 3.54^{+0.96}_{-0.85} $     &  $ 0.73^{+0.19}_{-0.18} $\\
	   \hline\hline
\end{tabular}
\end{center}
\end{table}
The errors arising from the uncertainty of the the continuum threshold values, the variation of the Borel mass parameter, etc., are taken into account. 
The magnetic moments of the the vector hidden-charmed tetraquark states have been extracted from the light-cone sum rules employing for their hadronic representations a single-pole technique [see, Eq. (\ref{edmn04})]. In the case of the multi-quark hadrons such technique should be verified by supplementary arguments because a physical representation of relevant sum rules receives contributions from two-hadron reducible terms as well. This problem was first proposed during theoretical studies of the pentaquarks \cite{Kondo:2004cr,Lee:2004xk}. Two-hadron contaminating terms have to be considered when extracting parameters of multi-quark hadrons. In the case of the multi-quark hadrons they lead to modification in the quark propagator
\begin{align}
\frac{1}{m^{2}-p^{2}} \rightarrow \frac{1}{m^{2}-p^{2}-i\sqrt{p^{2}}\Gamma (p)%
},  \label{eq:Modif}
\end{align}%
where $\Gamma (p)$ is the finite width of the  multi-quark hadrons generated by two-hadron scattering states.  When these effects are properly considered in the sum rules,  they rescale the residue of the  multi-quark hadrons under investigation leaving its mass unchanged. 
Detailed investigations show that two-hadron scattering effects are small  for  multi-quark hadrons (see Refs. \cite{Wang:2015nwa,Agaev:2018vag,Sundu:2018nxt,Wang:2019hyc,Albuquerque:2021tqd,Albuquerque:2020hio,Wang:2020iqt,Wang:2019igl,Wang:2020cme}).
Thus, in this study the zero-width single-pole approximation has been employed.

As we mentioned in the Introduction of the text, the same results are obtained for the spectroscopic parameters when the $[cq][\bar{c}\bar{q}]$, $[cs][\bar{c}\bar{s}]$  and $[cq][\bar{c}\bar {q'}]$ quark contents are taken into account. Therefore, the spectroscopic parameters are not enough to understand the internal structure of these states and to determine their quark contents.
Considering that the $Y_{c\bar c}$ states can have a $[cq][\bar{c}\bar{q}]$ or $[cs][\bar{c}\bar{s}]$ quark content, it is obvious that the magnetic moments of these states are zero. 
In this analysis, the $Y_{c\bar c}$ states are considered as quark content with $[cq][\bar{c}\bar {q'}]$ and their magnetic moments are obtained as nonzero.
These results provide a direct test of the diquark-antidiquark structure of the $Y_{c\bar c}$ states. In future experimental studies, the measurement of the magnetic moment will give us more detailed and clear information about the internal structure of these particles.

The magnitude of the magnetic moment shows its measurability in experiment. For instance, it shows that if the magnitude of the magnetic moment in the natural units ($\frac{ e}{2\, m_{Y_{c\bar c}}}$) is one or larger than it, it can be easily measured in the experiment. If the magnitude of the magnetic moment is less than one, it means that it is probable to be measured. We observe that the magnitudes of the magnetic moment results obtained in this study are large enough to be measured experimentally. 

To our best knowledge, this is the first study in the literature dedicated to the investigation of the $Y_{c\bar c}$ states magnetic moments. Therefore, experimental data or theoretical estimations are not yet available to compare them with our numerical results. However, we may compare these results with the $Z_c$ states’ magnetic moments.  Making this comparison may be meaningful in terms of having an idea about the consistency of the results since there is no experimental and theoretical data.
In Refs. \cite{Ozdem:2017jqh,Ozdem:2021yvo,Ozdem:2021hka}, the light-cone sum rules method has been applied various $Z_c$ states to obtain their electromagnetic properties.
 In Ref.\cite{Ozdem:2017jqh}, the electromagnetic properties of the tetraquark state $Z_c(3900)$ have been investigated in the diquark-antidiquark picture  with quantum numbers $J^{PC}=1^{+-}$ and its magnetic and quadrupole moments were extracted. The magnetic moment was obtained as $ \mu_{Z_c} =0.67 \pm 0.32 ~\mu_N$.
In Ref. \cite{Ozdem:2021yvo}, the magnetic dipole moment of the $Z_{cs}(3985)$ state was  acquired by using the molecular and compact diquark-antidiquark type interpolating currents. The obtained results were given as $\mu_{Z_{cs}}^{Di} =0.60^{+0.26}_{-0.21}~\mu_N$ and  $\mu_{Z_{cs}}^{Mol} =0.52^{+0.19}_{-0.17} ~\mu_N$ for diquark-antidiquark and molecular pictures, respectively.  
In Ref. \cite{Ozdem:2021hka}, the magnetic dipole moments of the $Z_{c}(4020)^+$, $Z_{c}(4200)^+$, $Z_{cs}(4000)^{+}$ and  $Z_{cs}(4220)^{+}$ states have been extracted using the hadronic molecular form of interpolating currents with quantum numbers $J^{PC}=1^{+-}$. The magnetic dipole moments were obtained as 
 $\mu_{Z_{c}} = 0.66^{+0.27}_{-0.25}~\mu_N$, 
 $\mu_{Z^{1}_{c}}=1.03^{+0.32}_{-0.29}~\mu_N$,
 $\mu_{Z_{cs}}=0.73^{+0.28}_{-0.26}~\mu_N$, and 
$\mu_{Z^1_{cs}}=0.77^{+0.27}_{-0.25}~\mu_N$ 
 for the  $Z_{c}(4020)^+$, $Z_{c}(4200)^+$, $Z_{cs}(4000)^{+}$ and  $Z_{cs}(4220)^{+}$ states, respectively.  As one can see from these predictions, the numerical results for the magnetic moments of the $Y_{c\bar c}$ states obtained in the present work are the same order of the $Z_c $ states' magnetic moments. 
Calculating the results for the magnetic moments with other theoretical models will also be a test of the consistency of our predictions.

Let us discuss how the magnetic moments of these states can be measured.
The electromagnetic multipole moments can be calculated using a method based on the emission of soft photons from hadrons, as presented in Ref.~\cite{Zakharov:1968fb}. The photon also contains information about the higher multipole moments of the particle when emitted. The element of the radiative process matrix can be written in terms of the energy of the photon  $E_\gamma$ as follows:
\begin{align}
 M \sim A \,(E_\gamma )^{-1} + B \,(E_\gamma )^0 + \mbox{higher-order terms}.
\end{align}
The electric charge contributes to the amplitude at order $(E_\gamma )^{-1}$ and the contribution coming from the magnetic  moment is characterized by the term $(E_\gamma)^0$. Thus, by measuring the cross section or decay width of the radiative process and ignoring  the small contributions of terms linear/higher order  in $E_\gamma$, one can identify the magnetic  moment of the state under examination.

In summary, the magnetic moments of the vector hidden-charmed tetraquark states that have been observed and can be expected to be observed experimentally have been determined using the light-cone sum rules taking into account the diquark-antidiquark structure with the quantum numbers $ J^{PC} =  1^{--}$  and $ J^{PC} =  1^{-+}$. 
Since these states are considered to have different flavors of light quarks, they have nonzero magnetic moments. The results obtained in this study can be checked for consistency by various methods. The magnetic moments of hadrons encompass useful knowledge about the distribution of charge and magnetization inside hadrons, which helps us to understand their geometrical shapes.
The existing theoretical estimations on the mass of vector hidden-charmed tetraquark states and their comparison with the experimental value have also led to different assignments on the internal structure of this state discussed above. More theoretical studies are needed, especially on the strong and radiative decays of these states. The values to be obtained can be very useful in terms of understanding the nature of these states when the results of this study are taken together.  Calculations of different parameters related to various interactions/decays of vector hidden-charmed tetraquark states and their comparison with likely future experimental measurements can help us figure out the substructure of these states.

  \begin{widetext}
  %
  \subsection*{Appendix: Explicit expression for \texorpdfstring{$\Delta_1 (M^2,s_0)$}{}}
 In this appendix we present the explicit expressions of the function $\Delta_1 (M^2,s_0)$ for the magnetic moments of $Y_{c\bar c}$ states entering into the sum rule.
  \begin{align}
   \Delta_1 (M^2,s_0) &= -\frac {3 (e_d - e_u+e_c)} {2621440 \pi^5}\Bigg[
   I[0, 6, 3, 0] - 4 I[0, 6, 3, 1] + 6 I[0, 6, 3, 2] - 
    4 I[0, 6, 3, 3] + I[0, 6, 3, 4] - 3 I[0, 6, 4, 0] \nonumber\\
    &+ 
    9 I[0, 6, 4, 1] - 9 I[0, 6, 4, 2] + 3 I[0, 6, 4, 3] + 
    3 I[0, 6, 5, 0] - 6 I[0, 6, 5, 1] + 3 I[0, 6, 5, 2] - 
    I[0, 6, 6, 0] \nonumber\\
    &+ I[0, 6, 6, 1] + 6 I[1, 5, 3, 1] - 
    18 I[1, 5, 3, 2] + 18 I[1, 5, 3, 3] - 6 I[1, 5, 3, 4] - 
    18 I[1, 5, 4, 1] + 36 I[1, 5, 4, 2] \nonumber\\
    &- 18 I[1, 5, 4, 3] + 
    18 I[1, 5, 5, 1] - 18 I[1, 5, 5, 2] - 6 I[1, 5, 6, 1]\Bigg]
    \nonumber\\
    &+ \frac {(e_d - e_u) m_c \langle g_s^2 G^2  \rangle \langle \bar q q \rangle } {442368 \pi^3}\Bigg[
   I[0, 2, 1, 0] - 2 I[0, 2, 1, 1] + I[0, 2, 1, 2] - 
    2 I[0, 2, 2, 0] + 2 I[0, 2, 2, 1]  \nonumber\\
    &+ I[0, 2, 3, 0]- 
    2 I[1, 1, 1, 0] + 4 I[1, 1, 1, 1] - 2 I[1, 1, 1, 2] + 
    4 I[1, 1, 2, 0] - 4 I[1, 1, 2, 1] - 2 I[1, 1, 3, 0]\Bigg]
        \nonumber\\
    &
    +\frac { m_c \langle g_s^2 G^2  \rangle \langle \bar q q \rangle  } {84934656 \pi^3}\Bigg[ 
   I[0, 2, 3, 
       0]\Bigg (3 e_d \Big (22 I_ 2[\mathcal{S}] - 
          24 I_ 2[\mathcal T_1] - 11 I_ 2[\mathcal T_2] + 
          1_3 I_ 2[\mathcal T_ 4] 
          - 2 I_ 2[\mathcal {\tilde S}] + 
          48 I_ 4[\mathcal T_1] \nonumber\\
    &
    + 22 I_ 4[\mathcal T_2] - 
          26 I_ 4[\mathcal T_4] - 16 I_ 5[\mathcal {A}]\Big) + 
       12 e_u \Big (22  I_ 1[\mathcal {S}] + 
           22 I_ 1[\mathcal T_1] - 11 I_ 1[\mathcal T_2] - 22 I_ 1[\mathcal T_4] + 68 I_1[\mathcal {\tilde S}] + 
           2 I_ 3[\mathcal T_1] 
           \nonumber\\
    &+ 11 I_ 3[\mathcal T_2] + 
           9 I_ 3[\mathcal T_4] - 92 I_ 3[\mathcal {\tilde S}] + 
           2 I_ 5[\mathcal {A}]\Big)\Bigg)
            - 
    48 e_d \mathcal {A}[u_ 0] \Bigg (I[0, 2, 1, 0] - 2 I[0, 2, 1, 1] +
        I[0, 2, 1, 2] 
        \nonumber\\
    &- 2 I[0, 2, 2, 0] + 2 I[0, 2, 2, 1] + 
       I[0, 2, 3, 0]\Bigg) + 
    8 \chi \Bigg (I_ 5[\varphi_ {\gamma}] \Big (8 e_d I[0, 3, 3, 0] - 
           4 e_u I[0, 3, 3, 0] - 3  e_d I[0, 3, 4, 0]
           \nonumber\\
    &- 
           3 e_u I[0, 3, 4, 0]\Big) + 
        2 e_d \Big (11 I[0, 3, 1, 0] - 33 I[0, 3, 1, 1] + 
           33 I[0, 3, 1, 2] - 11 I[0, 3, 1, 3] - 37 I[0, 3, 2, 0] 
           \nonumber\\
    &+ 
           74 I[0, 3, 2, 1] - 37 I[0, 3, 2, 2] + 41 I[0, 3, 3, 0] - 
           41 I[0, 3, 3, 1] - 15 I[0, 3, 4, 0]\Big) + 
        2 e_u \Big (-5 I[0, 3, 1, 0] 
        \nonumber\\
    &     + 19 I[0, 3, 1, 1] - 
            23 I[0, 3, 1, 2] + 9 I[0, 3, 1, 3] + 19 I[0, 3, 2, 0] - 
            46 I[0, 3, 2, 1] 
            \nonumber\\
    &+ 27 I[0, 3, 2, 2] - 23 I[0, 3, 3, 0] + 
            27 I[0, 3, 3, 1] + 9 I[0, 3, 4, 0]\Big) 
            \varphi_ {\gamma}[
           u_ 0]\Bigg)\Bigg]\nonumber\\
    &
      -\frac {f_ {3\gamma} \langle g_s^2 G^2  \rangle } {169869312 \pi^3}\Bigg[
   132 m_c^2\Big (4 e_u I_ 1[\mathcal V] - e_d I_ 2[\mathcal A] + 
       3 e_d I_ 2[\mathcal V]\Big) I[0, 2, 2, 0] + 
    384 e_d m_c^2 I_ 6[\psi^\nu] \big (2 I[0, 2, 2, 0] \nonumber\\
    &+ 
       3 I[0, 2, 3, 0]\big) - 
    16 (2 e_d + e_u) I_ 5[\psi^a] \big (24 m_c^2 I[0, 2, 2, 0] - 
       I[0, 3, 4, 0]\big) - 
    32 \Bigg (24 e_u m_c^2 \Big (I[0, 2, 1, 0] 
    \nonumber\\
    &- 
           I[0, 2, 2, 0]\Big) + 
        e_d \Big (24 m_c^2 \big (I[0, 2, 1, 0] - 2 I[0, 2, 1, 1] + 
               I[0, 2, 1, 2] - I[0, 2, 2, 0] + I[0, 2, 2, 1]\big) - 
            I[0, 3, 2, 0] 
            \nonumber\\
    &+ 2 I[0, 3, 2, 1] - I[0, 3, 2, 2] + 
            2 I[0, 3, 3, 0] - 2 I[0, 3, 3, 1] - 
            I[0, 3, 4, 0]\Big)\Bigg) \psi^a[u_ 0] + 
    192 m_c^2 \Bigg (2 e_u \Big (I[0, 2, 1, 1] 
    \nonumber\\
    &- I[0, 2, 1, 2] - 
           I[0, 2, 2, 1]\Big) + 
        e_d \Big (3 I[0, 2, 1, 0] - 8 I[0, 2, 1, 1] + 
            5 I[0, 2, 1, 2] - 6 I[0, 2, 2, 0] + 8 I[0, 2, 2, 1] 
            \nonumber\\
    &+ 
            3 I[0, 2, 3, 0]\Big)\Bigg) \psi^\nu[u_ 0]\bigg]     
    \nonumber\\
    &
    +\frac {(e_d - e_u)  \langle g_s^2 G^2  \rangle } {56623104 \pi^5}\Bigg[
   25 I[0, 4, 2, 0] - 75 I[0, 4, 2, 1] + 75 I[0, 4, 2, 2] - 
    25 I[0, 4, 2, 3] - 71 I[0, 4, 3, 0] \nonumber
           \end{align}
       \begin{align}
       &+ 142 I[0, 4, 3, 1] - 
    71 I[0, 4, 3, 2] + 67 I[0, 4, 4, 0] - 67 I[0, 4, 4, 1] - 
    21 I[0, 4, 5, 0] + 
    268 I[1, 3, 2, 1]\nonumber\\
     &  - 284 I[1, 3, 2, 2] + 100 I[1, 3, 2, 3] - 
    536 I[1, 3, 3, 1] + 284 I[1, 3, 3, 2] + 268 I[1, 3, 4, 1] + 64 m_c^2 \Big (2 I[0, 3, 1, 1]\nonumber\\
          &
     - 3 I[0, 3, 1, 2] + 
       I[0, 3, 1, 3] - 2 I[0, 3, 2, 1] + I[0, 3, 2, 2] + 
       3 I[1, 2, 1, 2] - 3 I[1, 2, 1, 3] +3 I[1, 2, 2, 2]\Big)\Bigg]
       \nonumber\\
    &
         +\frac {(e_d - e_u) m_c \langle \bar q q \rangle} {32768 \pi^3}\Bigg[
   I[0, 4, 2, 0] - 3 I[0, 4, 2, 1] + 3 I[0, 4, 2, 2] - 
    I[0, 4, 2, 3] - 3 I[0, 4, 3, 0] + 6 I[0, 4, 3, 1]  \nonumber\\
    &- 
    3 I[0, 4, 3, 2] + 3 I[0, 4, 4, 0] - 3 I[0, 4, 4, 1] - 
    I[0, 4, 5, 0] + 12 I[1, 3, 2, 1] - 12 I[1, 3, 2, 2] + 
    4 I[1, 3, 2, 3]  \nonumber\\
    & - 24 I[1, 3, 3, 1] + 12 I[1, 3, 3, 2] + 
    12 I[1, 3, 4, 1]\Bigg]\nonumber\\
    &- \frac {m_c \langle \bar q q \rangle} {5728640 \pi^3}\Bigg[-20 \Bigg (2 e_u \Big (6 I_1[\mathcal T_ 4] - 4 I_ 1[\mathcal {\tilde S}] + 5 I_ 3[\mathcal S] - 
          28 I_ 3[\mathcal T_ 1] - 24 I_ 3[\mathcal T_ 2] + 
          18 I_ 3[\mathcal T_ 3] + 10 I_ 3[\mathcal T_ 4] - 
          16 I_ 3[\mathcal {\tilde S}]\Big)
          \nonumber\\
    &+ 
       e_d \Big (-2 I_ 2[\mathcal S] + 2 I_ 2[\mathcal T_ 1] + 
           3 I_ 2[\mathcal T_ 3] + I_ 2[\mathcal T_ 4] + 
           2 I_ 2[\mathcal {\tilde S}] + 10 I_ 4[\mathcal S] - 
           16 I_ 4[\mathcal T_ 1] - 12 I_ 4[\mathcal T_ 2] + 
           4 I_ 4[\mathcal T_ 4] - 
           12 I_ 4[\mathcal {\tilde S}]\Big)\Bigg) I[0, 4, 4, 0] 
           \nonumber\\
    &+ 
    48 \chi \Bigg (-e_u \Big (2 I[0, 5, 2, 0] - 7 I[0, 5, 2, 1] + 
           9 I[0, 5, 2, 2] - 5 I[0, 5, 2, 3] + I[0, 5, 2, 4] - 
           6 I[0, 5, 3, 0] + 15 I[0, 5, 3, 1] 
           \nonumber\\
    &- 12 I[0, 5, 3, 2] + 
           3 I[0, 5, 3, 3] + 6 I[0, 5, 4, 0] - 9 I[0, 5, 4, 1] + 
           3 I[0, 5, 4, 2] - 2 I[0, 5, 5, 0] + I[0, 5, 5, 1]\Big) 
           \nonumber\\
    &+ 
        2 e_d \Big (I[0, 5, 2, 0] - 4 I[0, 5, 2, 1] + 
            6 I[0, 5, 2, 2] - 4 I[0, 5, 2, 3] + I[0, 5, 2, 4] - 
            3 I[0, 5, 3, 0] + 9 I[0, 5, 3, 1] - 9 I[0, 5, 3, 2] 
            \nonumber\\
    &+ 
            3 I[0, 5, 3, 3] + 3 I[0, 5, 4, 0] - 6 I[0, 5, 4, 1] + 
            3 I[0, 5, 4, 2] - I[0, 5, 5, 0] + 
            I[0, 5, 5, 1]\Big)\Bigg) \varphi_\gamma[u_ 0]\Bigg] \nonumber\\
            &+\frac {f_ {3\gamma}} {1966080 \pi^3}\Bigg[
   20 e_d m_c^2 \Big (I[0, 4, 2, 1] - 2 I[0, 4, 2, 2] + 
       I[0, 4, 2, 3] - 2 I[0, 4, 3, 1] + 2 I[0, 4, 3, 2] + 
       I[0, 4, 4, 1]\Big) 
       \nonumber\\
    &- 
    9 e_d \Big (I[0, 5, 3, 0] - 3 I[0, 5, 3, 1] + 3 I[0, 5, 3, 2] - 
       I[0, 5, 3, 3] - 3 I[0, 5, 4, 0] + 6 I[0, 5, 4, 1] - 
       3 I[0, 5, 4, 2] + 3 I[0, 5, 5, 0] 
       \nonumber\\
    &- 3 I[0, 5, 5, 1] - 
       I[0, 5, 6, 0]\Big) + 
    9 e_u \Big (I[0, 5, 3, 0] - 3 I[0, 5, 3, 1] + 3 I[0, 5, 3, 2] - 
        I[0, 5, 3, 3] - 3 I[0, 5, 4, 0] + 6 I[0, 5, 4, 1] 
        \nonumber\\
    &- 
        3 I[0, 5, 4, 2] + 3 I[0, 5, 5, 0] - 3 I[0, 5, 5, 1] - 
        I[0, 5, 6, 0]\Big)\Bigg] \psi^\nu[u_0],
           \end{align}
where $u_0= \frac{M_1^2}{M_1^2+M_2^2} $, $ \frac{1}{M^2}= \frac{1}{M_1^2}+\frac{1}{M_2^2}$ with $ M_1^2 $ and $ M_2^2 $ being the Borel parameters in the initial and final states, respectively. For simplicity we did not present the terms proportional to many higher dimensional operators here; however, in the numerical computations we take these terms into account. 
  
The ~$I[n,m,l,k]$ and $I_i[\mathcal{F}]$~functions are
defined as
\begin{align}
 I[n,m,l,k]&= \int_{4m_c^2}^{s_0} ds \int_{0}^1 dt \int_{0}^1 dw~ e^{-s/M^2}~
 s^n\,(s-4m_c^2)^m\,t^l\,w^k,\nonumber\\
 I_1[\mathcal{F}]&=\int D_{\alpha_i} \int_0^1 dv~ \mathcal{F}(\alpha_{\bar q},\alpha_q,\alpha_g)
 \delta'(\alpha_ q +\bar v \alpha_g-u_0),\nonumber\\
  I_2[\mathcal{F}]&=\int D_{\alpha_i} \int_0^1 dv~ \mathcal{F}(\alpha_{\bar q},\alpha_q,\alpha_g)
 \delta'(\alpha_{\bar q}+ v \alpha_g-u_0),\nonumber\\
    I_3[\mathcal{F}]&=\int D_{\alpha_i} \int_0^1 dv~ \mathcal{F}(\alpha_{\bar q},\alpha_q,\alpha_g)
 \delta(\alpha_ q +\bar v \alpha_g-u_0),\nonumber\\
   I_4[\mathcal{F}]&=\int D_{\alpha_i} \int_0^1 dv~ \mathcal{F}(\alpha_{\bar q},\alpha_q,\alpha_g)
 \delta(\alpha_{\bar q}+ v \alpha_g-u_0),\nonumber
   \end{align}
 \begin{align}
   I_5[\mathcal{F}]&=\int_0^1 du~ \mathcal{F}(u)\delta'(u-u_0),\nonumber\\
 I_6[\mathcal{F}]&=\int_0^1 du~ \mathcal{F}(u),\nonumber
 \end{align}
 where $\mathcal{F}$ stands for the corresponding photon DAs.

 \end{widetext}  
\bibliography{Yccbar-Diquark}

\end{document}